\documentclass{article}
\usepackage{spconf,amsmath,graphicx}
\usepackage{amssymb,amsfonts}
\usepackage[hyphens]{url}
\usepackage{breakurl}

\usepackage{bm,cite,mathtools}
\usepackage{subfig}
\usepackage{mathrsfs}

\usepackage{bbm}
\usepackage{float}
\usepackage{parselines}

\newcommand{\argmin}{\mathop{\rm arg~min}\limits}

\title{Region-to-region kernel interpolation of\\acoustic transfer function with directional weighting}
\name{Juliano G. C. Ribeiro, Shoichi Koyama, and Hiroshi Saruwatari}
\address{The University of Tokyo, 7-3-1 Hongo, Bunkyo-ku, Tokyo 113-8656, Japan}
\begin{document}
\ninept
\maketitle
\begin{abstract}
A method of interpolating the acoustic transfer function (ATF) between regions that takes into account both the physical properties of the ATF and the directionality of region configurations is proposed. Most spatial ATF interpolation methods are limited to estimation in the region of receivers. A kernel method for region-to-region ATF interpolation makes it possible to estimate the ATFs for both source and receiver regions from a discrete set of ATF measurements. We newly formulate the reproducing kernel Hilbert space and associated kernel function incorporating directional weight to enhance the interpolation accuracy. We also investigate hyperparameter optimization methods for this kernel function. Numerical experiments indicate that the proposed method outperforms the method without the use of directional weighting.
%We propose an acoustic transfer function (ATF) interpolation method that takes into account both the physical properties of the ATF and the directionality of the region configurations. Most well-established ATF representation models do not account for source position variation and rely on assumptions about the space analyzed, meaning these methods tend to be situational and impractical. The proposed method is based on the properties of the ATF alone. We incorporate the relative position of sources by defining a weighted inner-product. The associated hyperparameters are estimated using a leave-one-out cross validation criterion. The proposed method is compared to a previously established method with numerical simulations, where it is shown that the added directional weight improves kernel interpolation performance.
\end{abstract}
\section{Introduction}
\label{sec:intro}

The acoustic transfer function (ATF) characterizes sound propagation between two points in an acoustic environment, which is equivalent to the frequency response from a source to a receiver at these points. The ATF estimation of an environment has many applications, such as sound reproduction~\cite{Borra:ICASSP2019}, sound field equalization~\cite{Mazur:ICASSP2019}, echo cancellation~\cite{Pepe:EURASIP2010}, and speech dereverberation~\cite{Mohanan:ICASSP2017}.

%The acoustic transfer function (ATF) between two points relays frequency response of the impulse response (IR) between them. We consider propagation to be equivalent to a finite impulse response filter. There are many applications to understanding the ATF in a given environment, such as sound reproduction~\cite{Borra:ICASSP2019}, sound field equalization~\cite{Mazur:ICASSP2019}, echo cancellation~\cite{Pepe:EURASIP2010}, and speech dereverberation~\cite{Mohanan:ICASSP2017}.

The spatial interpolation of the ATF is widely studied because of its applicability. The ATF is generally represented as the frequency response of an impulse signal under the assumption of a linear time-invariant system. Thus, one method is to treat the ATF as a rational transfer function with poles/zeros~\cite{IIR_ATF,Haneda:IEEE_J_SAP1999}. 
Although the pole locations relative to the receiver position can be predicted, the relation between the locations of poles and/or zeros and the source position cannot. In addition, this method is dependent on the room shape and its eigenmodes.
%However, the relation between the location of poles and/or zeros and the source position is not known. In \cite{Haneda:IEEE_J_SAP1999}, the pole locations in  regards to receiver positions are known, but the method is dependent on the room's shape and acoustic properties. 
Several attempts have been made to spatially interpolate the ATF based on the sparsity of planewave components~\cite{Mignot:IEEE_ACM_J_ASLP2014, Antonello:TASLP2017}. However, those methods are limited to variable receiver positions within a receiver region, with a fixed source. 

%ATF interpolation is a vast area of study. One method is to treat the ATF as a standard rational transfer function with poles/zeros~\cite{Haneda:IEEE_J_SAP1999,IIR_ATF}. However, there is no explicit connection between poles/zeros and source/receiver positions. Methods that account for variable receivers do exist, such as~\cite{Mignot:IEEE_ACM_J_ASLP2014}, where the ATF is determined using compressed sensing, and~\cite{Antonello:TASLP2017}, which models the IR in regards to time and space. However, these methods do not account for variable source points.

Sound field estimation or reconstruction is a problem similar to the spatial interpolation of the ATF, and is aimed at estimating the continuous pressure distribution from a discrete set of microphones. Many sound field estimation methods are based on series expansions of finite-dimensional basis functions using planewaves and spherical wavefunctions~\cite{Poletti,Koyama:IEEE_J_ASLP2013}. An alternative approach is the \textit{kernel method}, where the solution space of the homogeneous Helmholtz equation is defined as the reproducing kernel Hilbert space (RKHS), and the estimate is obtained by kernel ridge regression~\cite{Ueno:IWAENC2018,Ueno:IEEE_SPL2018}. This infinite-dimensional analysis of the sound field makes it possible to estimate it without truncating the expansion order and has been used in various applications, such as spatial audio reproduction~\cite{Iijima:JASA_J_2021,Koyama:I3DA2021}, active noise cancellation~\cite{Koyama:IEEE_ACM_J_ASLP2021}, and sensor placement optimization~\cite{Ariga:ICASSP2020}. 

%The representation of the ATF spatially is largely dependent on the understanding of sound wave propagation, as well as sound fields. As solutions of the homogeneous Helmholtz equation, sound fields can be expressed as a series expansion of spherical wave functions regardless of boundary conditions~\cite{Poletti}. An alternative formulation is the use of kernel ridge regression~\cite{Ueno:IWAENC2018}.

In a prior work of the authors~\cite{Juliano:SAM2020}, an ATF interpolation method for both variable source and receiver positions, i.e., \textit{region-to-region ATF interpolation}, based on the kernel method was proposed. The definition of the RKHS is based on acoustic properties of ATFs: the constraint of satisfying the homogeneous Helmholtz equation for the reverberant component and acoustic reciprocity. This method significantly outperformed the method based on a finite-dimensional expansion of spherical wavefunctions proposed in~\cite{Samarasinghe}. 

%Using the sound field analysis techniques introduced in~\cite{Poletti}, a spatial ATF interpolation model was introduced in \cite{Samarasinghe}. The ATF is expressed as the superposition of a known direct component and a reverberant component that behaves similarly to a source-free sound field on both the source and receiver coordinates. Therefore, the reverberant component is represented as a spherical wave function series expansion. The ATF is thus approximated by a truncated series expansion added to the direct component.

In this paper, we extend the kernel method of region-to-region ATF interpolation to incorporate directional weight. The directional weight has recently been introduced in the kernel method for sound field estimation to take prior information on source directions into consideration~\cite{Ito:ICASSP2020,Ueno:IEEE_J_SP_2021}. We introduce the directional weight for the region-to-region ATF interpolation to incorporate knowledge of the configuration of source and receiver regions. Since the use of directional weight requires the determination of several hyperparameters of the reproducing kernel functions, we also investigate hyperparameter optimization methods. Numerical experiments are conducted to compare the proposed method with the method without directional weighting in order to evaluate the effect of directionality on the estimations.

%In~\cite{Juliano:SAM2020}, a similar framework is used and the ATF is estimated by kernel ridge regression (KRR) constrained by the properties of the ATF. This method also included a property of the ATF not accounted for in~\cite{Samarasinghe}, the acoustic reciprocity. Both methods are capable of interpolating the ATF for variable source and receiver positions within assigned regions. However, the KRR outperformed the previous method in interpolating the ATF within the assigned regions.

%We now expand the idea of the KRR in~\cite{Juliano:SAM2020} by inserting directional weighting into the inner-product formulation in a similar manner to~\cite{Ito:ICASSP2020}. However, the introduction directionality also introduced hyperparameters that must be determined. We compare the proposed method with the previously established KRR method in numerical simulations.

\section{Problem statement and preliminaries}
\label{sec:krr}

\begin{figure}[!t]
\centering
\centerline{\includegraphics[width=0.8\columnwidth]{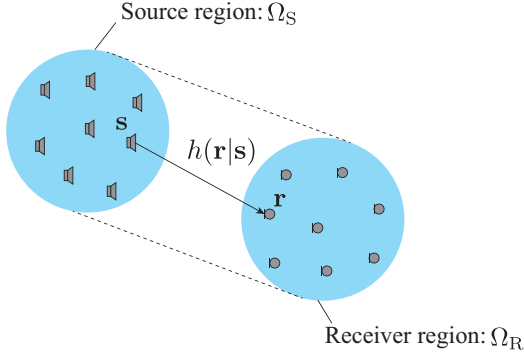}}
  \caption{Schematic diagram of a region-to-region ATF interpolation problem.}
  \label{fig:prob}
\end{figure}

Given a space $\Omega \subset \mathbb{R}^3$ with stationary acoustic properties, our objective is to estimate the ATF $h \colon \Omega_{\mathrm{R}} \times \Omega_{\mathrm{S}} \to \mathbb{C}$ between any source/receiver pair of positions $\mathbf{r}\in \Omega_\mathrm{R}\subset \Omega$ and $\mathbf{s}\in \Omega_\mathrm{S}\subset \Omega$, where $\Omega_\mathrm{S}$ and $\Omega_\mathrm{R}$ are the source and receiver regions, respectively (see Fig.~\ref{fig:prob}). 

\subsection{Preliminaries}

We assume that any ATF $h$ can be separated into two components: a known direct component $h_\mathrm{D}$ given as the free-field Green's function $G_0$ and an unknown reverberant component $h_\mathrm{R}$, as shown in \cite{Samarasinghe,Juliano:SAM2020}. $h_\mathrm{R}$ satisfies the homogeneous Helmholtz equation~\cite{EGWilliams:FourierAcoustics}. These assumptions are represented as
%In~\cite{Juliano:SAM2020}, it was shown that any ATF function $h$ can be separated into two components: a known direct component $h_\mathrm{D}$ and an unknown reverberant component $h_\mathrm{R}$ that is known to fulfill the Helmholtz equation~\cite{EGWilliams:FourierAcoustics}.
\begin{align}
&h(\mathbf{r}|\mathbf{s},k) = h_\mathrm{D}(\mathbf{r}|\mathbf{s},k)+h_\mathrm{R}(\mathbf{r}|\mathbf{s},k)\\
&h_\mathrm{D}(\mathbf{r}|\mathbf{s},k) = G_0(\mathbf{r}|\mathbf{s},k) = \frac{e^{\mathrm{i}k\|\mathbf{r}-\mathbf{s}\|}}{4\pi \|\mathbf{r}-\mathbf{s}\|}\\
&(\nabla^2_\mathbf{r}+k^2) h_\mathrm{R}(\mathbf{r}|\mathbf{s}, k) =(\nabla^2_\mathbf{s}+k^2) h_\mathrm{R}(\mathbf{r}|\mathbf{s}, k) = 0,
\end{align}
where $\mathbf{r}|\mathbf{s}$ is the source/receiver pair of positions, $\mathrm{i}$ is the imaginary unit, $\nabla^2_\mathbf{r}$ and $\nabla^2_\mathbf{s}$ are the Laplacian operators on the coordinates of $\mathbf{r}$ and $\mathbf{s}$, respectively, $k=2\pi \breve{f} /c$ is the wavenumber, $\breve{f}$ is the frequency, and $c$ is the speed of sound. Hereafter, $k$ in the argument of the ATFs is omitted for notational simplicity.

\subsection{Region-to-region ATF interpolation problem}

We distribute a set of $M$ receivers at points $\{\mathbf{r}_m\}_{m=1}^M$ and $L$ sources at points $\{\mathbf{s}_l\}_{l=1}^L$ to obtain a total of $N$ ($=LM$) ATF measurement values. We collectively denote  $\mathbf{q}_{n}=\mathbf{r}_m|\mathbf{s}_l$ for the position pairs with index $n=m+(l-1)M$ ($\in\{1, \ldots, N\}$). A set of $N$ ATFs is given, and then the direct component is removed from them to obtain the measurement vector $\mathbf{y}=[y_1, \ldots, y_N]^\mathsf{T}$ corresponding to each $\mathbf{q}_n$. We define our optimization problem as
\begin{align}
& \hat{h}_{\mathrm{R}} = \argmin_{f\in\mathscr{H}} \mathcal{J}(f) \notag\\
& \mathcal{J}(f) \coloneqq \sum_{n=1}^N |y_n-f(\mathbf{q}_n)|^2+\lambda \|f\|_{\mathscr{H}}^2,\ f\in \mathscr{H},
\label{eq:optprob}
\end{align}
where $\lambda>0$ is the regularization constant and $\mathscr{H}$ is the feature space to which the interpolation of the reverberant component belongs. 
%We then optimize $\mathcal{J}$ in regards to the feature space $\mathscr{H}$: 
%\begin{equation}
%\hat{h}_\mathrm{R} = \underset{f\in \mathscr{H}}{\mathrm{argmin}} \left ( \mathcal{J} \right )
%\end{equation}
This interpolation function is then added to the direct component to obtain $\hat{h}$:
\begin{equation}
\hat{h}(\mathbf{r}|\mathbf{s}) = h_\mathrm{D}(\mathbf{r}|\mathbf{s})+\hat{h}_\mathrm{R}(\mathbf{r}|\mathbf{s}).
\end{equation}

\subsection{Kernel ridge regression}

We consider the space $\mathscr{H}$ to be a reproducing kernel Hilbert space (RKHS). That is, it is a Hilbert space $(\mathscr{H}, \langle \cdot, \cdot \rangle_\mathscr{H})$ that admits a reproducing kernel function $\kappa$. The reproducing kernel is a bivariate function that satisfies:
\begin{equation}
\langle \kappa(\cdot, \mathbf{r}|\mathbf{s}) , f\rangle_\mathscr{H} = f(\mathbf{r}|\mathbf{s}),\ \forall \mathbf{r}\in \Omega_\mathrm{R},\ \forall \mathbf{s}\in \Omega_\mathrm{S}.
\end{equation}
In that case, the interpolation function $\hat{h}_\mathrm{R}$ is given by kernel ridge regression: 
\begin{equation}
\hat{h}_\mathrm{R}(\mathbf{r}|\mathbf{s}) = \bm{\kappa} (\mathbf{r}|\mathbf{s}) (\mathbf{K}+\lambda \mathbf{I})^{-1}\mathbf{y},
\label{eq:krr}
\end{equation}
where $\bm{\kappa}(\mathbf{r}|\mathbf{s}) = [ \kappa(\mathbf{r}|\mathbf{s}, \mathbf{q}_1), \dots, \kappa(\mathbf{r}|\mathbf{s}, \mathbf{q}_N)]$ is the kernel function vector, $\mathbf{I}$ is the identity matrix, and $\mathbf{K}$ is the Gram matrix defined as
\begin{equation}
\mathbf{K} = 
\begin{bmatrix} 
\kappa(\mathbf{q}_1, \mathbf{q}_1) & \kappa(\mathbf{q}_1, \mathbf{q}_2) & \dots & \kappa(\mathbf{q}_1, \mathbf{q}_N) \\
 \vdots           &            \vdots       & \ddots & \vdots \\
 \kappa(\mathbf{q}_N, \mathbf{q}_1) & \kappa(\mathbf{q}_N, \mathbf{q}_2) & \dots & \kappa(\mathbf{q}_N, \mathbf{q}_N)
\end{bmatrix}.
\end{equation}

\section{Region-to-region ATF interpolation with directional weighting}

%In \cite{Ito:ICASSP2019}, the authors define the directional weight for the sound field problem. The addition of directionality to the model shows notable improvements in its interpolation prowess.

We previously formulated an RKHS for solving \eqref{eq:optprob} based on the spherical wavefunction expansion in a previous study~\cite{Juliano:SAM2020}. We now formulate an RKHS based on the planewave expansion incorporating directional weight to take into consideration the behavior of the ATF with respect to the relative positions of $\Omega_\mathrm{R}$ and $\Omega_\mathrm{S}$.

%We defined a RKHS formulation for the ATF problem also embedded with directional weighting using similar reasoning. The weight is defined taking into consideration the behaviour of the ATF in regards to the relative positions of $\Omega_\mathrm{R}$ and $\Omega_\mathrm{S}$.

\subsection{Feature space definition}

Since the reverberant component $h_{\mathrm{R}}$ does not include any sources in $\Omega_{\mathrm{S}}$ and $\Omega_{\mathrm{R}}$, $h_{\mathrm{R}}$ can be represented by a planewave expansion, i.e., \textit{Herglotz wavefunction}~\cite{Colton2003}, as
%In order to define a RKHS for the reverberant component, we will make use of the Herglotz wave function~\cite{Ikehata:HMJ2005}:
\begin{align}
 & h_\mathrm{R}(\mathbf{r}|\mathbf{s}) = \mathcal{I}\left(\tilde{h}_\mathrm{R};\mathbf{r}|\mathbf{s}\right),\\
 & \mathcal{I}\left(f ; \mathbf{r}|\mathbf{s} \right) \coloneqq \int_{\mathbb{S}^2 \times \mathbb{S}^2} e^{\mathrm{i}k(\hat{\mathbf{r}}\cdot \mathbf{r}+\hat{\mathbf{s}}\cdot \mathbf{s})} f(\hat{\mathbf{r}},\hat{\mathbf{s}})\mathrm{d}\hat{\mathbf{r}}\mathrm{d}\hat{\mathbf{s}}, 
\label{eq:Herglotz}
\end{align}
where $\mathcal{I}$ is the operator for the planewave expansion, and $\mathbb{S}^2$ is the set of vectors in $\mathbb{R}^3$ of unit norm, representing directions. 
%The plane wave expansion represents the expression of the transmitted signal as the superposition of several plane waves.
Therefore, $\tilde{h}_\mathrm{R}(\hat{\mathbf{r}}, \hat{\mathbf{s}})$ is the complex amplitude of the planewave component of $h_\mathrm{R}$ 
from the source direction $\hat{\mathbf{s}}$ to the receiver direction $\hat{\mathbf{r}}$.
%that leaves the source at a direction $\hat{\mathbf{s}}$ and, after being reflected and transformed by the environment, arrives at the receiver through the direction $\hat{\mathbf{r}}$. 
The reciprocity of $h_\mathrm{R}$ imposes the condition that $\tilde{h}_\mathrm{R}(\hat{\mathbf{r}}, \hat{\mathbf{s}})=\tilde{h}_\mathrm{R}(\hat{\mathbf{s}}, \hat{\mathbf{r}})$. Thus, our inner-product space $(\mathscr{H}, \langle \cdot , \cdot \rangle_\mathscr{H})$ is defined as
\begin{equation}
\begin{split}\mathscr{H} = & \left \{ h_\mathrm{R} = \mathcal{I}\left(\tilde{h}_\mathrm{R};\mathbf{r}|\mathbf{s}\right) \colon \tilde{h}_\mathrm{R}\in L^2(W,\mathbb{S}^2\times \mathbb{S}^2),\right.\\
&\left. \hspace{70pt}  \tilde{h}_\mathrm{R}(\hat{\mathbf{r}}, \hat{\mathbf{s}})=\tilde{h}_\mathrm{R}(\hat{\mathbf{s}}, \hat{\mathbf{r}}) \ \forall \hat{\mathbf{r}}, \hat{\mathbf{s}}\in \mathbb{S}^2   \vphantom{\hat{h}} \right \} \end{split} 
\end{equation}
\begin{equation}
\langle f,g\rangle_\mathscr{H} = \int_{\mathbb{S}^2\times \mathbb{S}^2} \frac{\overline{\tilde{f}(\hat{\mathbf{r}}, \hat{\mathbf{s}})} \tilde{g}(\hat{\mathbf{r}}, \hat{\mathbf{s}})}{W(\hat{\mathbf{r}}, \hat{\mathbf{s}})} \mathrm{d}\hat{\mathbf{r}} \mathrm{d}\hat{\mathbf{s}},\ \forall f,g\in \mathscr{H},
\end{equation}
where $\overline{\ \cdot\ }$ is the complex conjugate, $W \colon \mathbb{S}^2 \times \mathbb{S}^2 \to \mathbb{R}_+$ is a directional weighting function, and $L^2(W,\mathbb{S}^2\times \mathbb{S}^2)$ is the space of functions of bounded square integral for $W$. Under these conditions, $(\mathscr{H}, \langle \cdot , \cdot \rangle_\mathscr{H})$ inherits the completeness of $L^2$ spaces~\cite{Rudin:FunctionalAnalysis} and as such is a Hilbert space. We can show that it is an RKHS by showing that the following function is its reproducing kernel:

To show that this function is the reproducing kernel, we only have to show that it satisfies the projection property for any given $f\in \mathscr{H}$:
\begin{align}
\langle \kappa(\cdot , \mathbf{r}|\mathbf{s}), f\rangle_\mathscr{H} %\notag \\
%&\ \ \ \ \ \ \ \ \ \ \ \ \ \ \ \ \ \ 
&= \int_{\mathbb{S}^2\times \mathbb{S}^2}  \frac{e^{\mathrm{i}k(\hat{\mathbf{r}}\cdot \mathbf{r}+\hat{\mathbf{s}}\cdot \mathbf{s})}\tilde{f}+ e^{\mathrm{i}k(\hat{\mathbf{r}}\cdot\mathbf{s}+\hat{\mathbf{s}}\cdot \mathbf{r})} \tilde{f}}{2}\mathrm{d}\hat{\mathbf{r}}\mathrm{d}\hat{\mathbf{s}}\notag \\
%&\ \ \ \ \ \ \ \ \ \ \ \ \ \ \ \ \ \ 
&=\frac{1}{2}\left (\mathcal{I}\left (\tilde{f};\mathbf{r}|\mathbf{s}\right) + \mathcal{I}\left (\tilde{f};\mathbf{s}|\mathbf{r}\right) \right )\notag \\
&=\frac{f(\mathbf{r}|\mathbf{s})+f(\mathbf{s}|\mathbf{r})}{2}.
\end{align} 
Since $f$ is reciprocal, we have
\begin{equation}
\langle \kappa(\cdot , \mathbf{r}|\mathbf{s}), f\rangle_\mathscr{H} = f(\mathbf{r}|\mathbf{s}).
\end{equation}

We also consider that the directional weighting function $W$ is separable for $\hat{\mathbf{r}}$ and $\hat{\mathbf{s}}$, that is,
%Showing that indeed the space is a RKHS. Another consideration made was that the weight would be separable. That is:
\begin{equation}
W(\hat{\mathbf{r}},\hat{\mathbf{s}} ) = w(\hat{\mathbf{r}})w(\hat{\mathbf{s}}),
\end{equation}
where $w \colon \mathbb{S}^2 \to \mathbb{R}_+$ is a directional weighting function in a single direction.

\subsection{Proposed directional weighting function} 

Since the direct component is removed from the measurements to obtain the reverberant component, the directional weighting function for the region-to-region ATF interpolation should have minimal gain in the direction connecting the centers of both regions, which is defined as $\hat{\mathbf{v}}_0$. Therefore, we define the directional weighting function $w$ as
 \begin{equation}
 w(\hat{\mathbf{v}}) = \frac{1}{4\pi}\left (1+\gamma^2-\frac{\cosh (\beta \hat{\mathbf{v}} \cdot \hat{\mathbf{v}}_0)}{\cosh (\beta)}\right ),\ \hat{\mathbf{v}}\in \mathbb{S}^2.
 \label{eq:dir_weight}
 \end{equation}
The hyperparameters $\beta$ and $\gamma$ characterize $\Omega$ acoustically. Since the direct components are expected to be weaker, the early reflections 
%(represented by the lateral plane wave components) 
are expected to be significantly more influential. An increase in $\beta$ makes the weight more selective in regards to the lateral components, while $\gamma$ sets the minimum gain baseline for all directions.
%  Here, the hyperparameter $\beta$ represents the level of selectivity of the method: the greater $\beta$ is, the more selective the kernel is with respect to directions neighboring $\hat{\mathbf{v}}_0$. The hyperparameter $\gamma$ adjusts the gain between the direct and lateral components.
 
\begin{figure}[!t]
\centering
\centerline{\includegraphics[width=0.70\columnwidth]{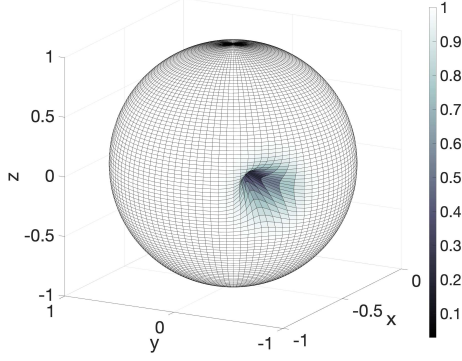}}
  \caption{Directional weighting function $w$ when $\hat{\mathbf{v}}_0 = [ -1/\sqrt{2}, -1/\sqrt{2}, 0]^\mathsf{T}$, $\gamma = 0.01$, and $\beta = 100$. The gain is given by the distance from the origin to the surface, quantified according to the bar on the right.}
  \label{fig:weight}
\end{figure}

 To better illustrate this, an example of the directional weighting function is plotted in Fig.~\ref{fig:weight}, which shows that the weight is minimal in the chosen direction $\hat{\mathbf{v}}_0$ but maximal in the lateral directions. $\beta$ controls the width of the cavity passing through $\hat{\mathbf{v}}_0$, while $\gamma$ controls the depth.
\subsection{Relation to prior work using uniform weight}

By setting $\gamma =1$ and $\beta=0$, the directional weighting function becomes uniform, i.e., $w = 1/4\pi$. The reproducing kernel of this uniform weight has the closed form
%For the second weight, we considered $w\equiv 1/4\pi$, equivalent to $\gamma =1$ and $\beta=0$. For this case, the RK of this uniform weight has a closed form:
\begin{equation}
\begin{split}
\kappa(\mathbf{r}|\mathbf{s}, \mathbf{r}^\prime|\mathbf{s}^\prime) = \frac{1}{2} \left (j_0(k\|\mathbf{r}-\mathbf{r}^\prime\|)j_0(k\|\mathbf{s}-\mathbf{s}^\prime\|) \right. \\
\ \ \ \ \left. +j_0(k\|\mathbf{s}-\mathbf{r}^\prime\|)j_0(k\|\mathbf{r}-\mathbf{s}^\prime\|) \right ),
\end{split}
\end{equation}
where $j_0$ is the $0$th-order spherical Bessel function of the first kind. This kernel function is identical to that used in \cite{Juliano:SAM2020}, making both estimations equivalent.
\section{Hyperparameter optimization for directional weight}
\label{sec:hyper}

The introduction of hyperparameters in \eqref{eq:dir_weight} also necessitates a criterion for choosing them. Although the profile of the reverberant component is understood, the exact balance between the direct and lateral component gains is not known outright.
There are several methods of hyperparameter optimization used for kernel ridge regression~\cite{Horiuchi:WASPAA2021,Nozal:JASA2021}. For this application, we employ leave-one-out cross-validation (LOO) because of its simplicity and nearly unbiased nature~\cite{Pontil:IJSC2002}.

We begin the computation of LOO by selecting a data point and measurement point pair $(\mathbf{q}_n, y_n)$ to remove from the data set. We then derive the desired model using the remaining $N-1$ elements of the data set and compute the error in estimating $y_n$ with this partial model. The value of LOO is the average of all errors calculated by repeating this process exhaustively:
\begin{equation}
\mathrm{LOO(\mathbf{y}, \ell)} = \frac{1}{N} \sum_{n=1}^N \ell \left(\hat{f}_n(\mathbf{q}_n)-y_n\right ),
\end{equation}
where $\ell$ is the chosen loss function, and $\hat{f}_n$ is the model derived when considering all data pairs except $(\mathbf{q}_n, y_n)$. We consider two types of loss function $\ell$: square error (SQE) and Tukey loss.
\begin{align}
&\mathrm{SQE}(z) = |z|^2 \\
&\mathrm{Tukey}(z) = 
\begin{cases} 
\displaystyle \frac{\sigma^2}{6} \left (1- \left ( 1-\frac{|z|^2}{\sigma^2}\right )^3\right ),\ |z|\leq \sigma \\ 
\displaystyle \frac{\sigma^2}{6},\ |z|> \sigma   \end{cases}, 
\label{eq:loss_tukey}
\end{align}
where $z$ is a complex variable, and $\sigma$ is the selectivity parameter. Tukey loss is more selective than SQE, which means that these loss functions have similar behavior near $0$, but Tukey loss has slower growth for higher discrepancies. This selectivity makes Tukey loss more resilient to outliers.

Both SQE and Tukey loss are differentiable with respect to the hyperparameters, meaning that LOO can be optimized using gradient descent methods. We applied the improved robust back-propagation algorithm introduced in~\cite{Anastasiadis:proceedings2003}.

\section{Numerical simulations}

\label{sec:num_sim}

\begin{figure}[!t]
\centering
\centerline{\includegraphics[width=0.8\columnwidth]{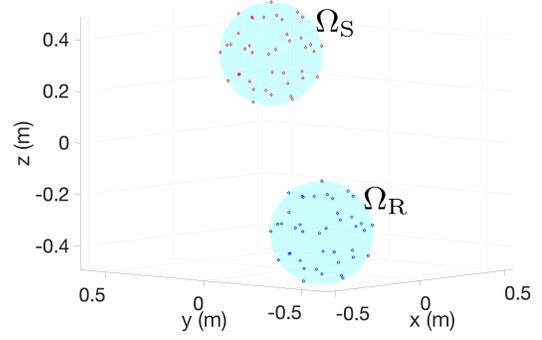}}
  \caption{Experimental setup utilized for the simulations. Red diamonds represent sources and blue circles represent receivers.}
  \label{fig:exp_setup}
\end{figure}

The proposed method using directional weight was evaluated using 3D acoustic simulations based on the image source method~\cite{Allen:JASA1979} by comparing it with the method using uniform weight. The room used for the evaluation was shoebox-shaped and had $3.2~\mathrm{m} \times 4.0~\mathrm{m} \times  2.7~\mathrm{m}$ dimensions. The reflection coefficients of the walls were set so that the reverberation time $T_{60}$ was $0.45~\mathrm{s}$. The source and receiver regions were spheres of $0.2~\mathrm{m}$ radius whose centers were $\mathbf{s}_0 = [ 0.35, 0.43, 0.29]^\mathsf{T}~\mathrm{m}$ for $\Omega_\mathrm{S}$ and $\mathbf{r}_0 = -\mathbf{s}_0$ for $\Omega_\mathrm{R}$. The origin of the coordinate system was set at the center of the room. We used $\hat{\mathbf{v}}_0 = (\mathbf{r}_0-\mathbf{s}_0)/\|\mathbf{r}_0-\mathbf{s}_0\|$ for the directional kernel.

%The simulation parameters were set for a reverberation time of $T_{60}=0.45~\mathrm{s}$, signal-noise ratio of $20~\mathrm{dB}$ and room dimensions $3.2~\mathrm{m} \times 4.0~\mathrm{m} \times  2.7~\mathrm{m}$. 
The measurement points were given for a total of $L=M=41$ points distributed on two spherical layers. The point distribution was given by the spherical $t$-design~\cite{Chen:Spherical}: $t=4$ for the outer layer and $t=3$ for the inner layer. We also added Gaussian noise so that the signal-to-noise ratio was $20~\mathrm{dB}$. The parameter $\sigma$ in the Tukey loss \eqref{eq:loss_tukey} was set to $0.4$, which was determined so that the probability of the error falling within four standard deviations was above $99\%$. The regularization parameter $\lambda$ in \eqref{eq:krr} was set to $10^{-2}$.
%Considering the noise distribution to follow a Gaussian, the probability of the error factoring within four standard deviations is over $99\%$. We considered this selectivity to be sufficient and thus chose $\sigma = 4.10^{-\mathrm{SNR}/20}=0.4$. The regularization constant was $\lambda = 10^{-\mathrm{SNR}/10}=10^{-2}$.

The evaluation measure was defined as the normalized mean square error (NMSE),
%We compared both methods using the normalized mean square error (NMSE) evaluated for frequencies ranging from $100$ to $1150~\mathrm{Hz}$ on a total of $N^\prime = 9025$ source/receiver pairs $\mathbf{q}^\prime$. The NMSE was calculated as:
\begin{equation}
\mathrm{NMSE} = 10 \log_{10} \left ( \frac{\sum_n \left |\hat{h}(\mathbf{q}^\prime_n) - h(\mathbf{q}^\prime_n)\right |^2}{\sum_n |h(\mathbf{q}^\prime_n)|^2}\right ),
\end{equation}
where $\mathbf{q}^\prime_n$ denote the $n$th evaluation source/receiver pair. We set  $9025$ source/receiver pairs of evaluation. 

\begin{figure}[!t]
\centering
\centerline{\includegraphics[width=1.0\columnwidth]{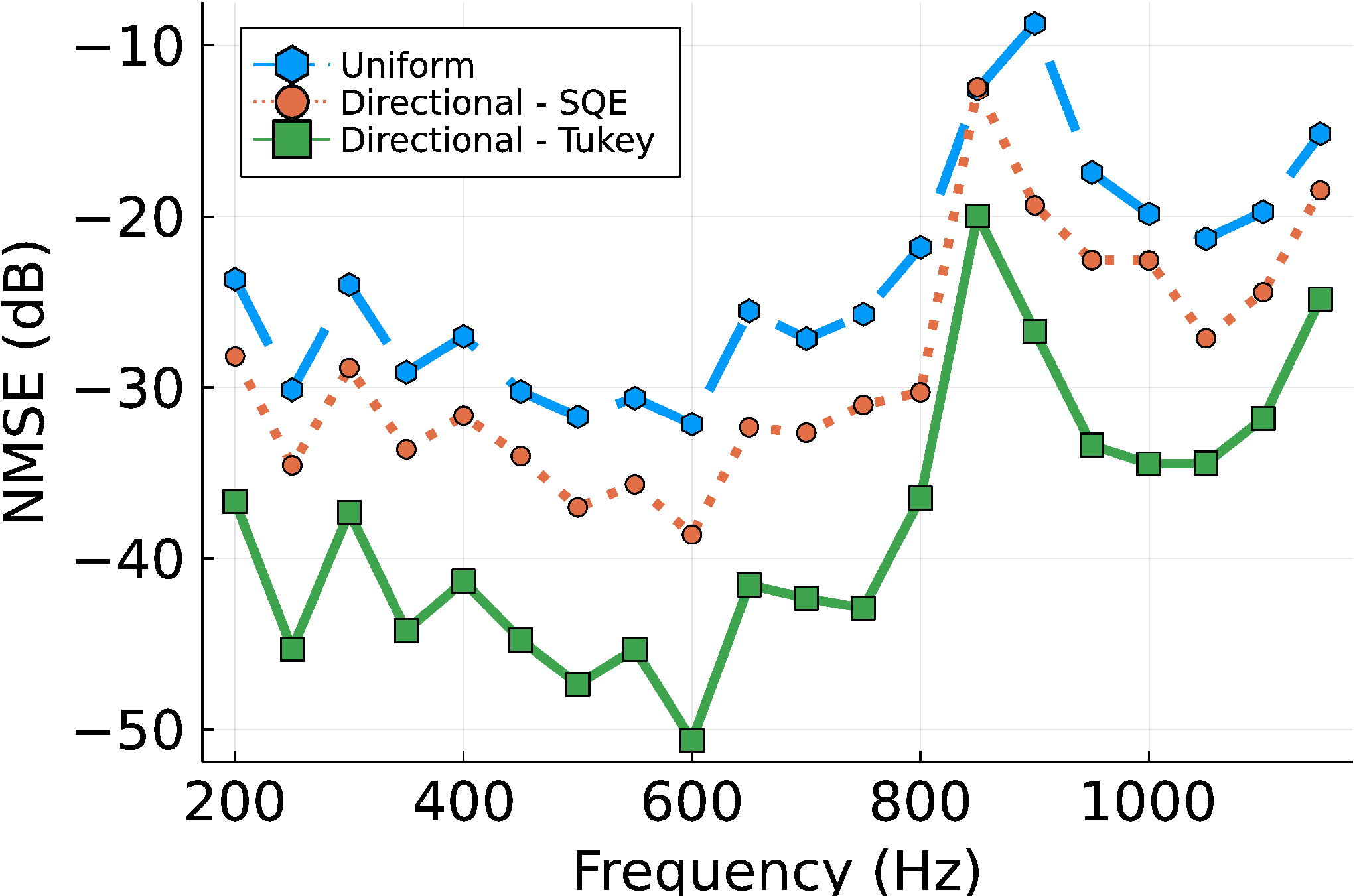}}
  \caption{NMSE performance of the methods compared.}
  \label{fig:NMSE}
\end{figure}

\begin{figure}[!t]
\centering
 \subfloat[True]{\includegraphics[width=0.42\columnwidth]{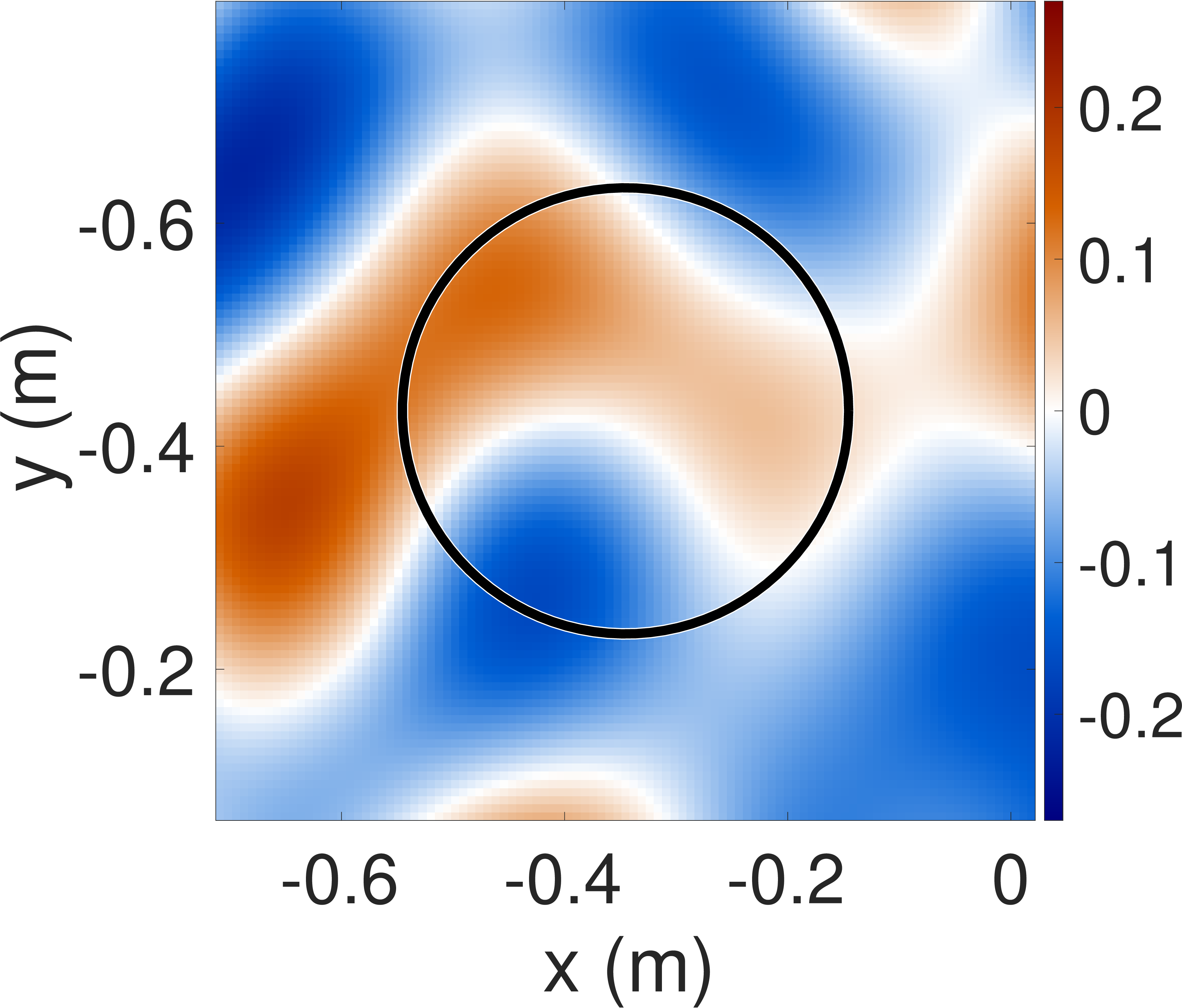}}
  \subfloat[Uniform]{\includegraphics[width=0.42\columnwidth]{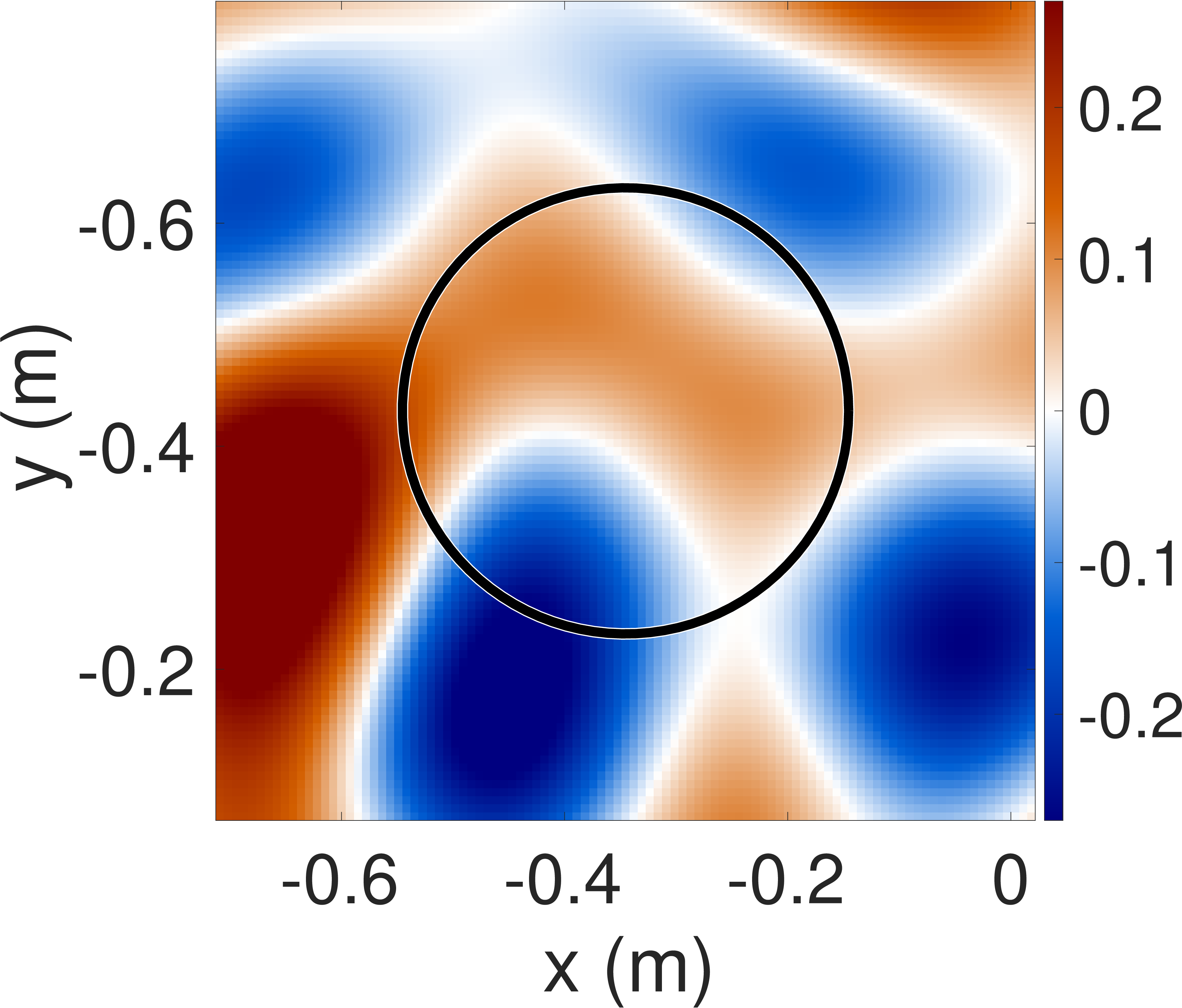}}
  \\
  \centering
   \subfloat[Directional - SQE]{\includegraphics[width=0.42\columnwidth]{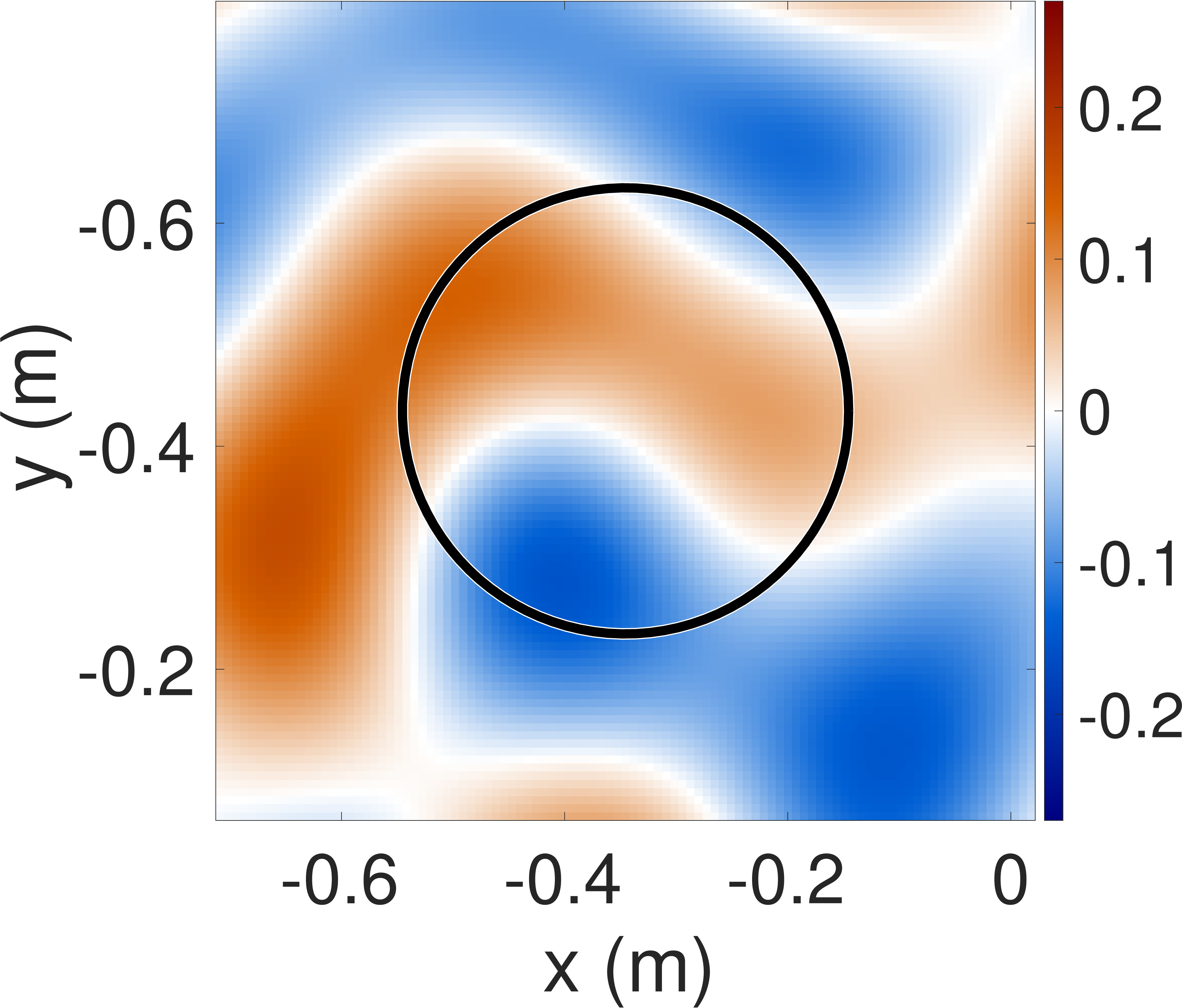}}
     \subfloat[Directional - Tukey]{\includegraphics[width=0.42\columnwidth]{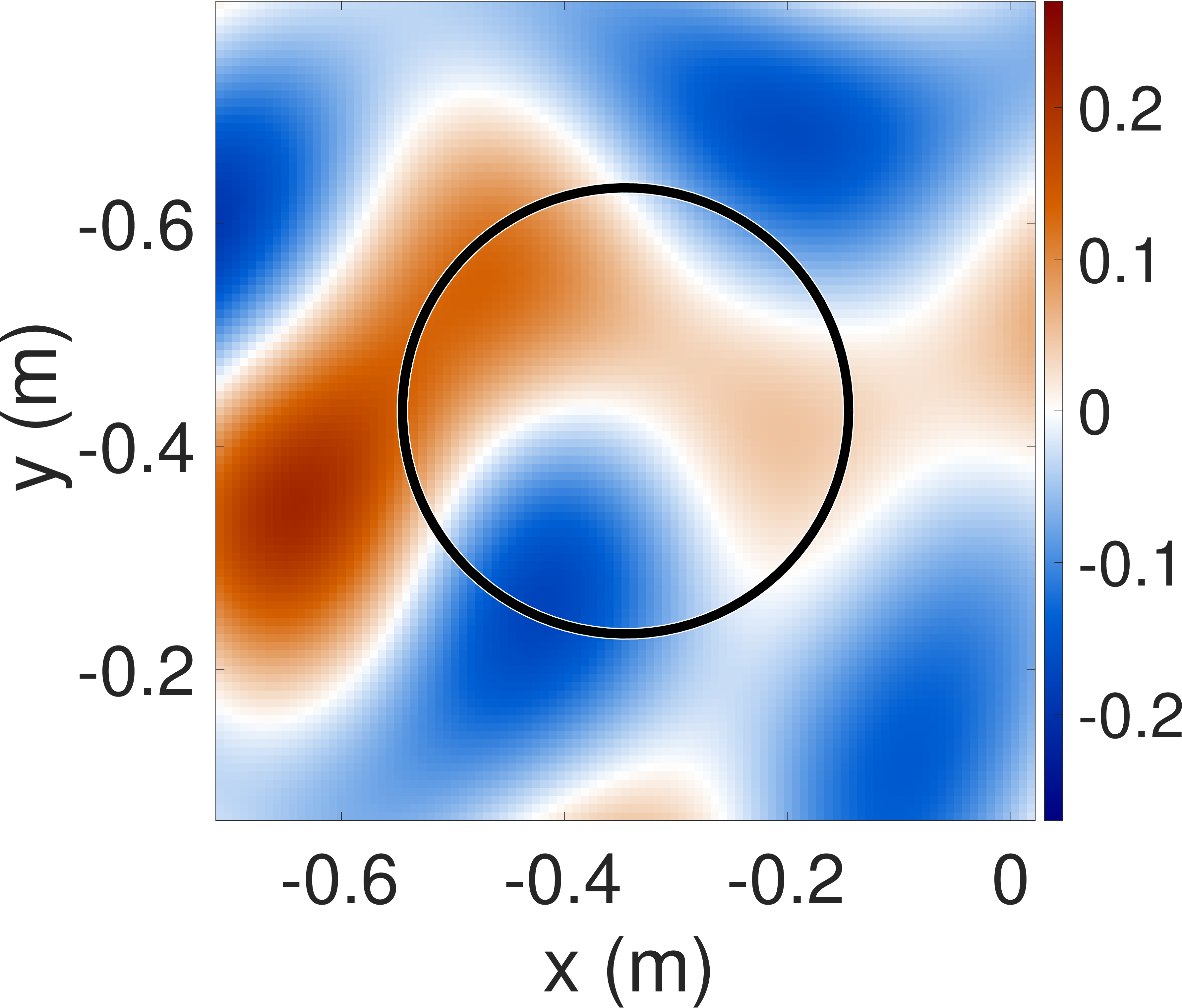}}
  \caption{Distributions of true and estimated ATFs in $\Omega_{\mathrm{R}}$ from the center of $\Omega_\mathrm{S}$ at $[0.35 , 0.43 , 0.29]^\mathsf{T}~\mathrm{m}$ for $950~\mathrm{Hz}$. The black circle indicates the bounds of $\Omega_\mathrm{R}$.}
  \label{fig:reconst}
 \end{figure}

\begin{figure}[!t]
\centering
  \subfloat[Uniform]{\includegraphics[width=0.42\columnwidth]{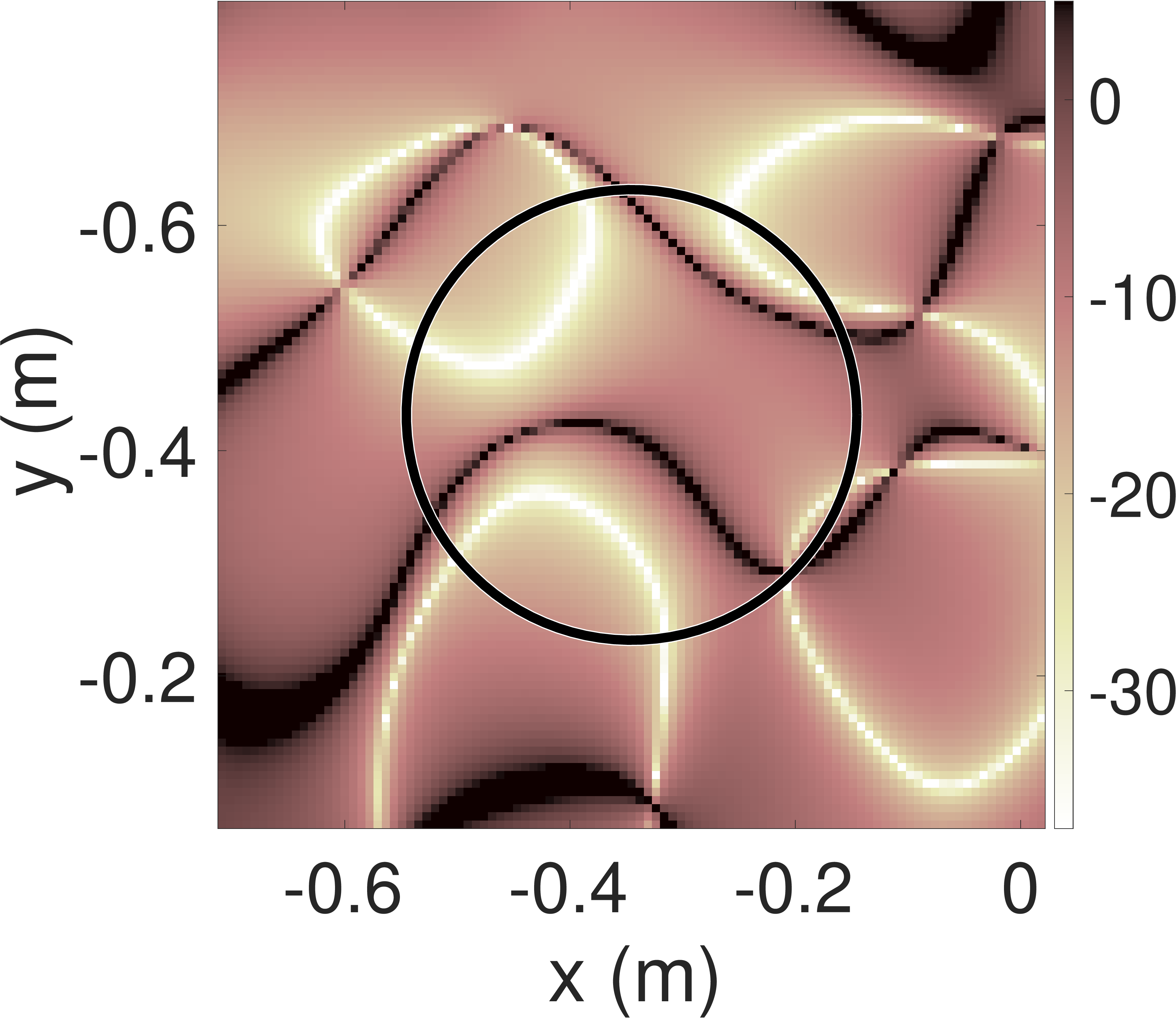}}
  \\
  \centering
  \subfloat[Directional - SQE]{\includegraphics[width=0.42\columnwidth]{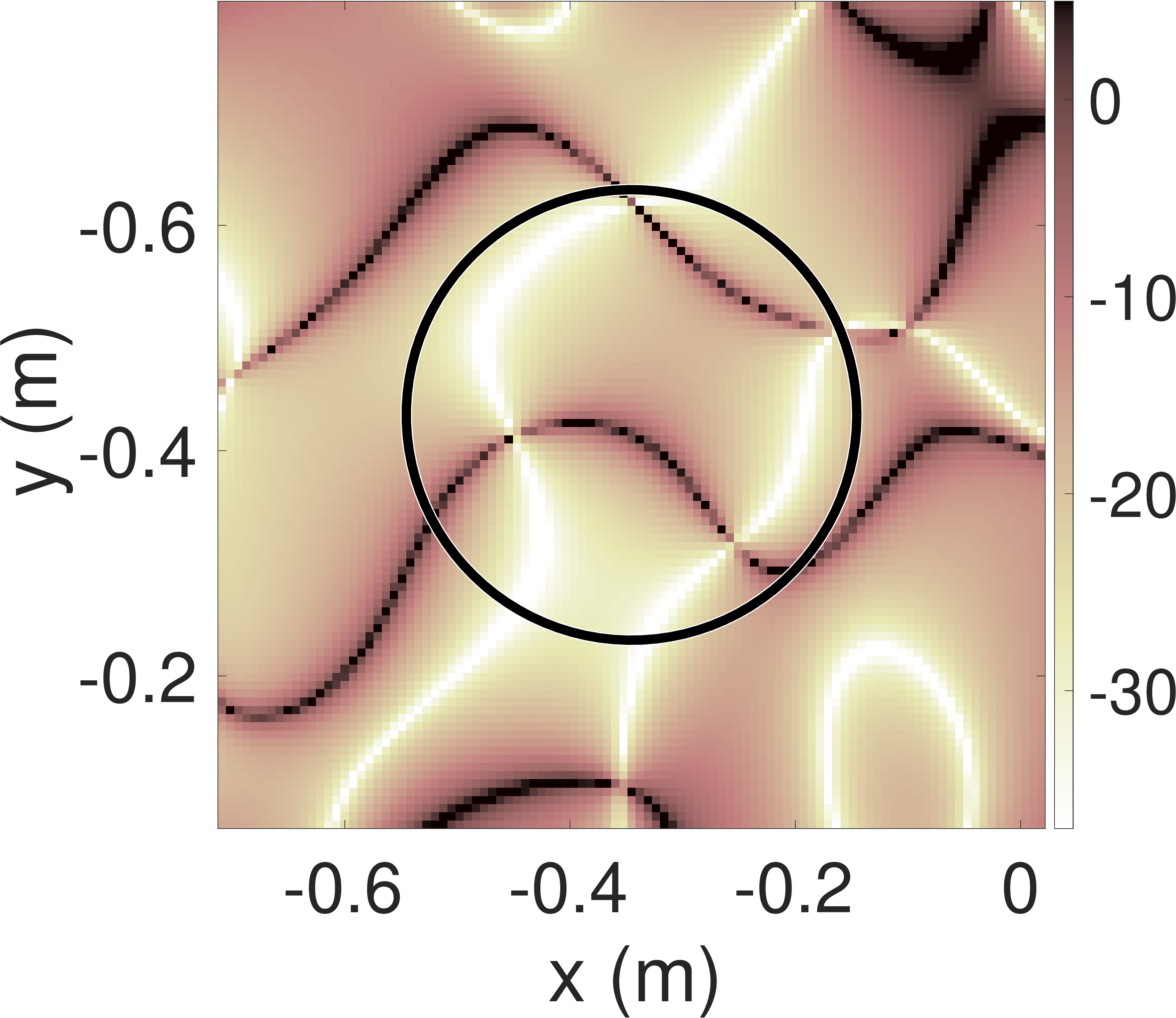}}
  \subfloat[Directional - Tukey]{\includegraphics[width=0.42\columnwidth]{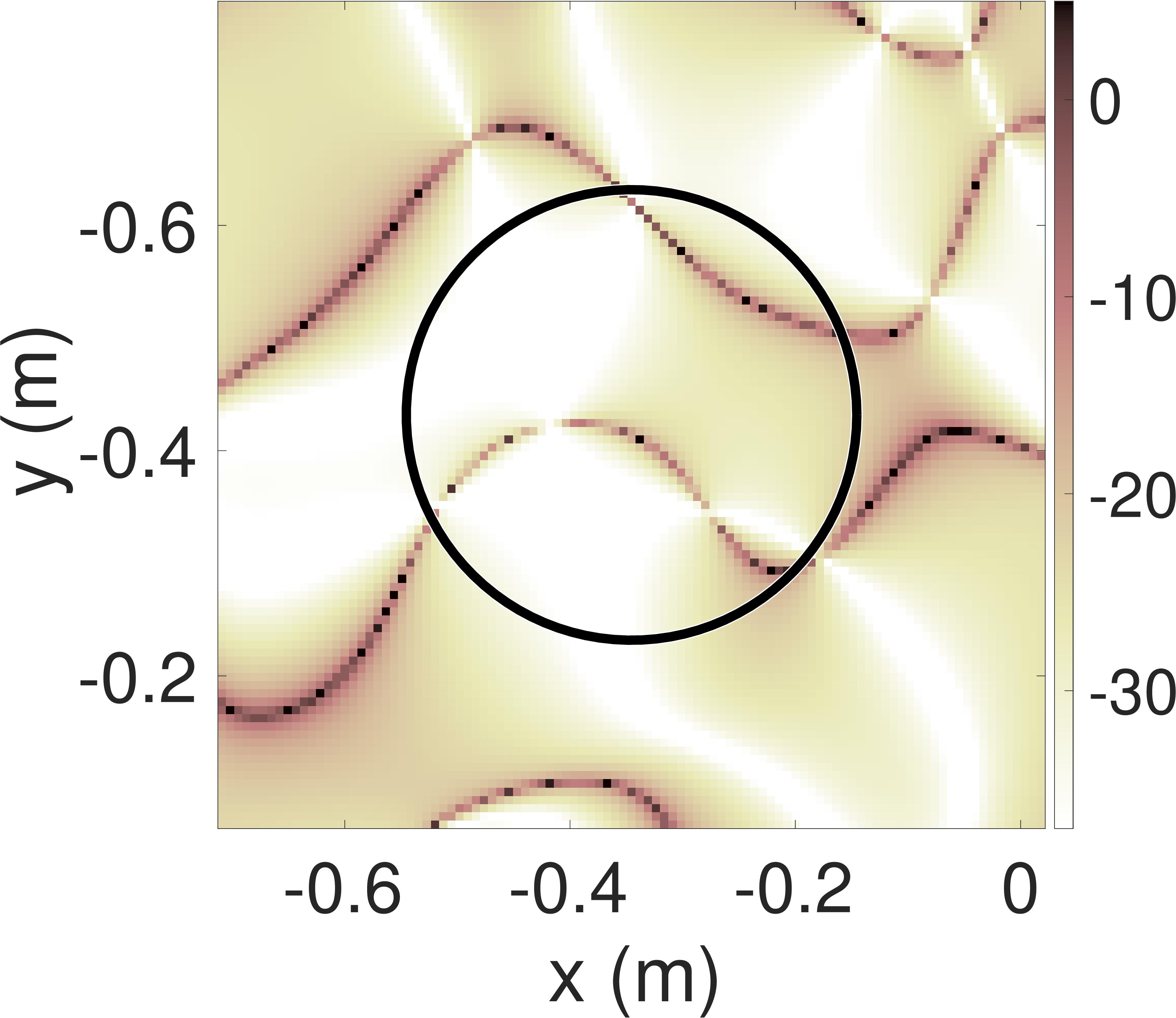}}
  \caption{Distributions of NSEs in $\Omega_{\mathrm{R}}$.}
  \label{fig:NSE}
\end{figure}

The NMSE with respect to the frequency ranging from $100$ to $1150~\mathrm{Hz}$ is shown in Fig.~\ref{fig:NMSE}. We also show the estimated ATF in $\Omega_\mathrm{R}$ generated by a single source point at $[ 0.35, 0.43, 0.29]^\mathsf{T}~\mathrm{m}$ for $950~\mathrm{Hz}$ in Fig.~\ref{fig:reconst}. The normalized squared error distribution of this plot is also shown in Fig.~\ref{fig:NSE}, which is defined as
%These results can be seen in Fig.~\ref{fig:NMSE}. We also showcase the reconstruction of a sound field in $\Omega_\mathrm{R}$ generated by a single source point at $\mathbf{s}_0$ in Fig.~\ref{fig:reconst}. Another criterion of analysis was the normalized square error (NSE) for a frequency of $950~\mathrm{Hz}$. These results are shown in Fig.~\ref{fig:NSE}.
\begin{equation}
\mathrm{NSE}(\mathbf{r}) = 10\log_{10} \left ( \frac{\left |h(\mathbf{r}|\mathbf{s}_0)-\hat{h}(\mathbf{r}|\mathbf{s}_0)\right |^2}{|h(\mathbf{r}|\mathbf{s}_0)|^2}\right ).
\end{equation}
The proposed method using directionally weighted kernels achieved a lower NMSE than the uniform counterpart for every frequency. For the SNR of $20~\mathrm{dB}$, the more selective Tukey loss also significantly outperformed the simpler SQE used in LOO. These results indicate that the directional weight is effective for the region-to-region ATF interpolation problem. In addition, LOO with Tukey loss is more useful for hyperparameter optimization than the simple SQE loss.

%The optimization of the hyperparameters lowered the NMSE for every frequency significantly. This shows LOO was successful in estimating weight parameters that better represented the behavior of the ATF. The application of Tukey loss resulted in further improvements and an estimation more resilient to noise.

%The optimized directional weights were also able to better reconstruct the ATF in a pointwise manner. Not only are NSE values generally lower for the proposed formulations, the areas where the methods are considered reliable are also larger. Among the proposed methods, the Tukey loss-based estimation displayed more robustness than the SQE once again.

\section{Conclusion}
\label{sec:conc}

We proposed a kernel interpolation method for region-to-region ATF interpolation with directional weighting. The reproducing kernel Hilbert space and associated reproducing kernel function are formulated on the basis of planewave decomposition, having properties of the Helmholtz equation constraint, acoustic reciprocity, and directionality. The spatial interpolation of the ATF is achieved by kernel ridge regression using this kernel function. Hyperparameters included in the kernel function are optimized by leave-one-out cross-validation. In the numerical experiments, the proposed method achieved highly accurate interpolation compared with the method using uniform weight. Furthermore, robustness to the noise was significantly improved by using Tukey loss in LOO.

%We proposed a kernel interpolation method of the ATF using a kernel function determined with directional weighting. The hyperparameters associated with the directional weighting were determined using a cross-validation criterion, LOO.

%For the hyperparameter estimation, LOO was shown to be unbiased in regards to the dataset, therefore not leading to overfitting. The proposed directionally-weighted kernel was shown to outperform the previously established uniform kernel methodology in numerical simulations. The proposed methods outperformed the uniform kernel both in a frequency and point-by-point basis, showing the optimization of the weights also leads to better region-to-region performance.

%Furthermore, for the relatively low SNR, the more selective Tukey loss was able to outperform the square loss. Therefore, we conclude it is possible to select hyperparameters that improve the noise robustness of the KRR.

\section{Acknowledgements}
This work was supported by JSPS KAKENHI Grant Number JP19H01116 and JST PRESTO Grant Number JPMJPR18J4.

\vfill\pagebreak

\bibliographystyle{IEEEbib_mod}
\bibliography{str_def_abrv,koyama_en,refs}

\end{document}